\documentclass[a4paper,11pt]{article}
\usepackage{pos}
\usepackage{aas_macros}

\DeclareRobustCommand{\ion}[2]{%
	\relax\ifmmode
	\ifx\testbx\f@series
	{\mathbf{#1\,\mathsc{#2}}}\else
	{\mathrm{#1\,\mathsc{#2}}}\fi
	\else\textup{#1\,{\mdseries\textsc{#2}}}%
	\fi}

\title{
Confirmation of the Star J004229.87+410551.8 in M31 As a B[e] Supergiant
}

\author*[a]{Sarkisyan A.}
\author[a]{Vinokurov A.}
\author[a]{Solovyeva Yu.}
\author[b]{Atapin K.}
\author[a]{Sholukhova O.}
\author[a]{Fabrika S.}
\author[c]{Bizyaev D.}

\affiliation[a]{Special Astrophysical Observatory of the Russian Academy of Sciences,\\ Nizhny~Arkhyz, Karachai-Cherkessia, Russia}

\affiliation[b]{Sternberg Astronomical Institute, Moscow M. V. Lomonosov State University,\\ Universitetskiy prospekt~13, Moscow, Russia}

\affiliation[c]{Apache Point Observatory and New Mexico State University,\\
	 2001 Apache Point Road, Sunspot, NM, USA}

\emailAdd{ars@sao.ru}

\abstract{

We study the luminous blue variable candidate J004229.87+410551.8 in the Andromeda Galaxy.
Earlier, the star displayed a spectral anomaly: although a hot emission spectrum had been detected, it had strong \ion{Ca}{ii} H and K  absorption lines. Subsequently the star was assumed to be a hot hypergiant or a B[e] supergiant. For the purpose of clear star classification, we conducted spectroscopic and photometric analysis of the object and its surroundings with the 6-m telescope of SAO RAS, the 3.5-m ARC telescope of the Apache Point Observatory and the 2.5-m telescope of the Caucasus Mountain Observatory of the Sternberg Astronomical Institute. The spectrum of the star has the \ion{Fe}{ii}, [\ion{Fe}{ii}], [\ion{O}{i}], and Balmer emission lines. Its spectral energy distribution shows an excess in the near-infrared range due to hot circumstellar dust. The indicated features and a high estimated value of star’s luminosity ($\log (L/L_{\odot}) = 4.6\pm 0.2$) allow us to finally classify the object as a B[e] supergiant.

}

\FullConference{%
  The Multifaceted Universe: Theory and Observations -  2022 (MUTO2022)\\
  23-27 May 2022\\
  SAO RAS, Nizhny Arkhyz, Russia\\}

\begin{document}
\maketitle

\section{Introduction}

The star J004229.87+410551.8 was selected for this study from the sample of luminous blue variable
candidates \cite{Massey2007}. The authors note the peculiarity of its spectrum: although the Balmer lines are in emission and the lines of [\ion{Fe}{ii}] are present, the strength of the \ion{Ca}{ii} H and K lines and the presence of the G band could indicate a much cooler absorption spectrum. A suggestion was made that the spectrum would be composite. Subsequently Humphreys et al.~\cite{Humphreys2014} did not find absorption features of F or G stars in the obtained spectrum, and the star was assumed to be a hot hypergiant~\cite{Humphreys2013} or a B[e]-supergiant \cite{Humphreys2017a}. The goal of this work is to clarify the star's classification on the basis of additional spectroscopic and photometric analysis. We will refer to the star by the first part of its identifier in the Local Group Galaxies Survey (LGGS) \cite{Massey2006}.

\section{Observations and archival data}

\begin{figure}
	\begin{minipage}{0.49\textwidth}
		\centering
		\includegraphics[height=2.75cm]{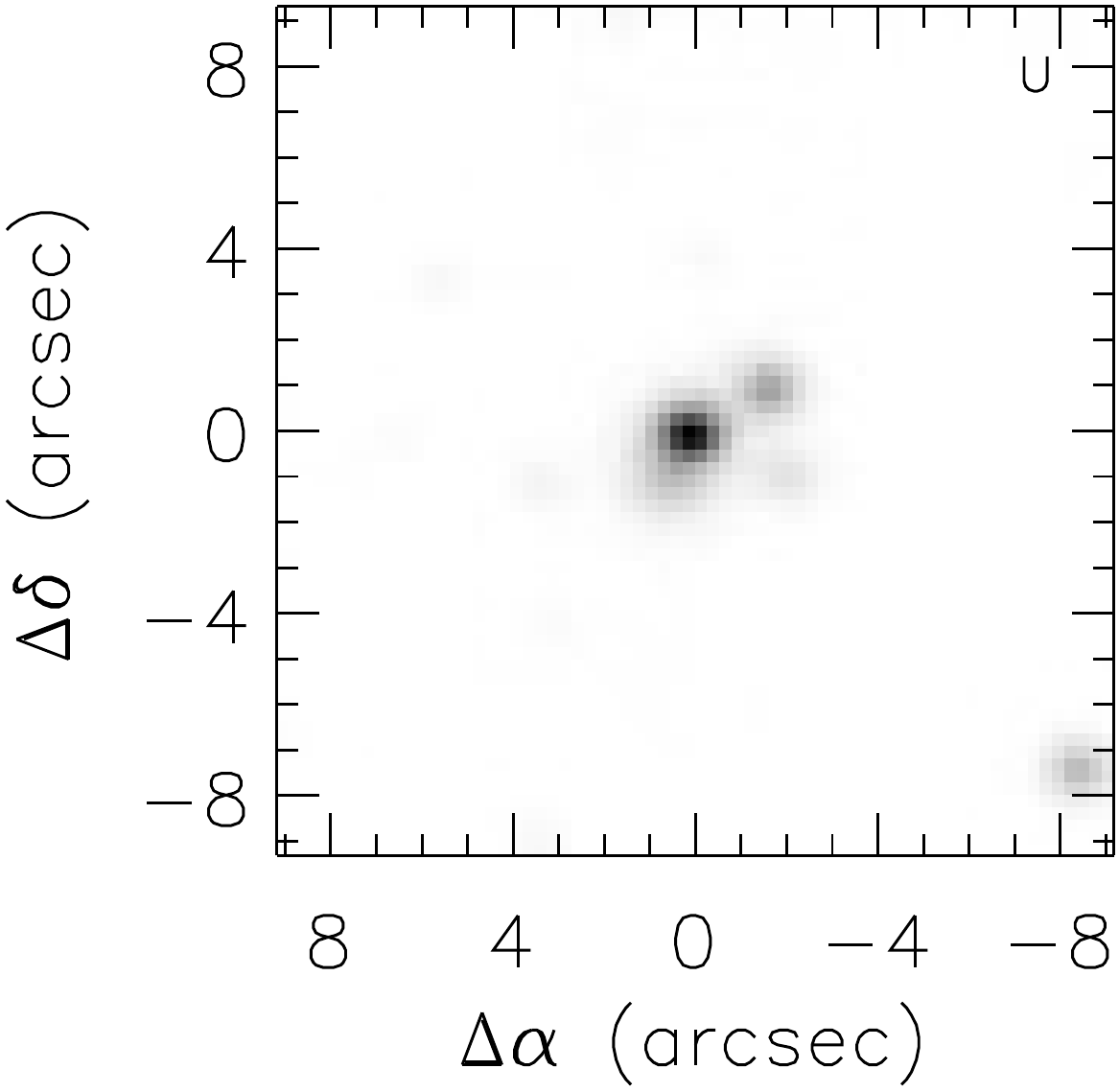}
		\includegraphics[trim=80 0 0 0,clip,height=2.75cm]{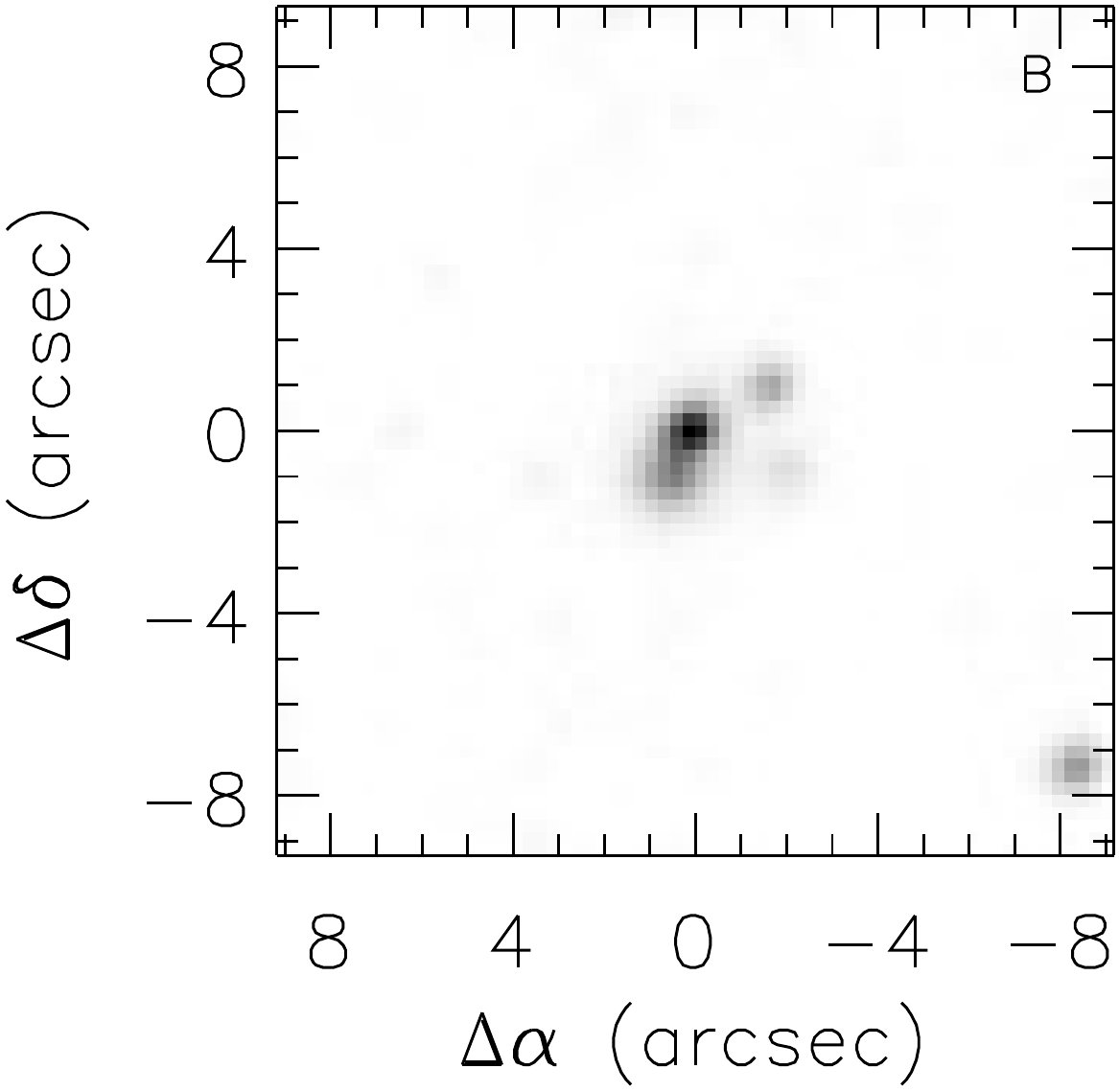}
		\includegraphics[trim=80 0 0 0,clip,height=2.75cm]{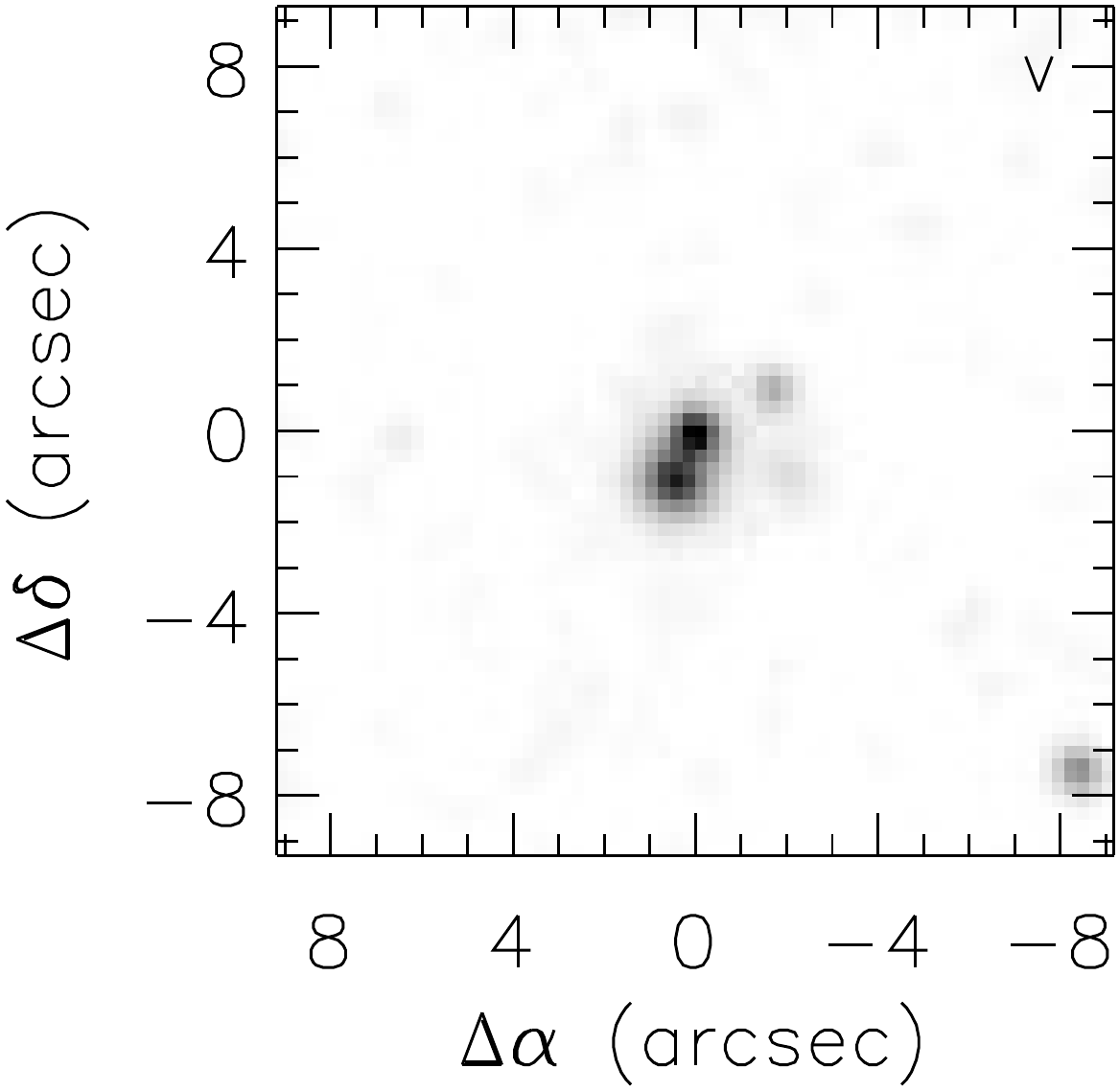}
		\includegraphics[height=2.75cm]{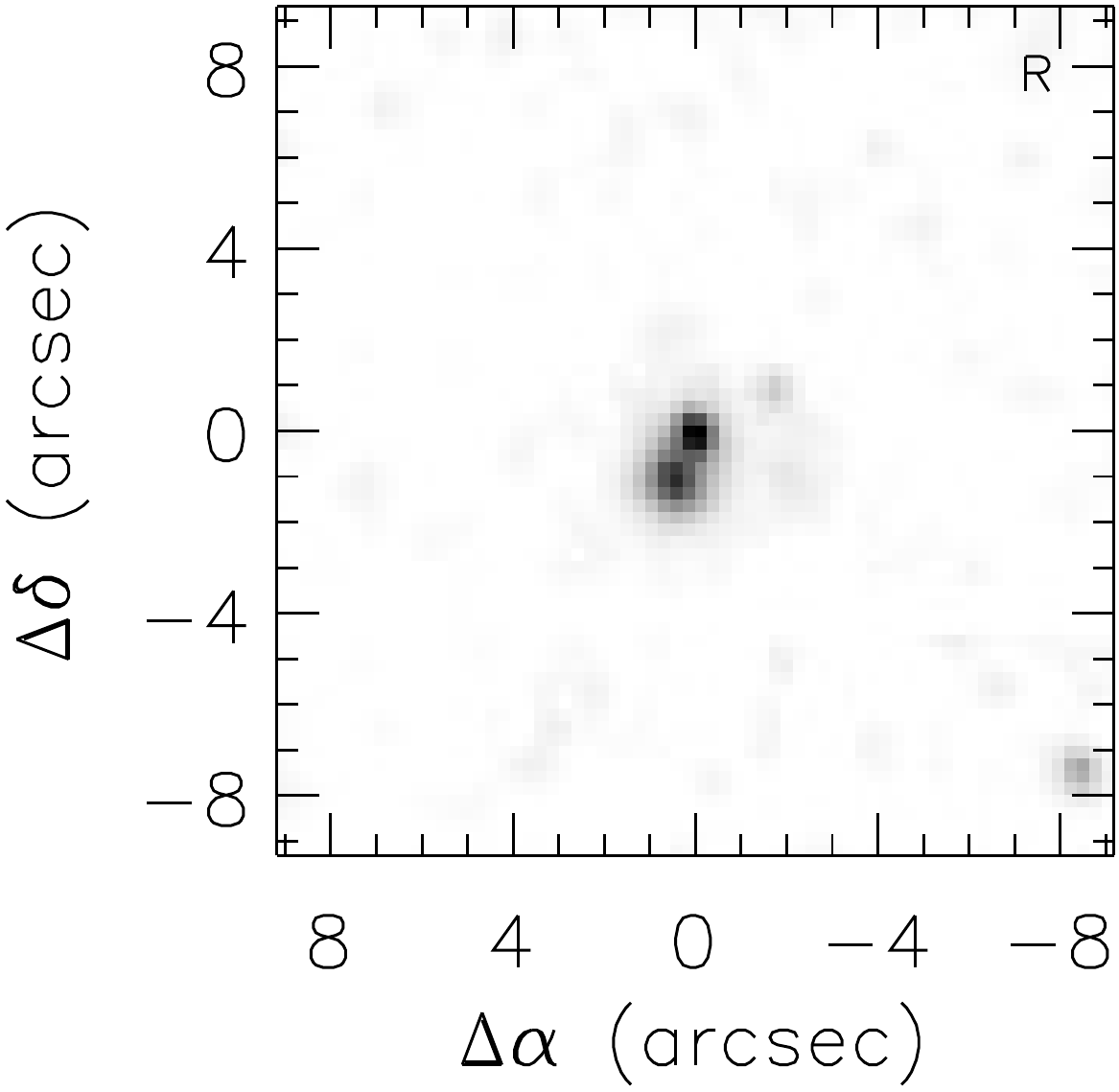}	
		\includegraphics[trim=80 0 0 0,clip,height=2.75cm]{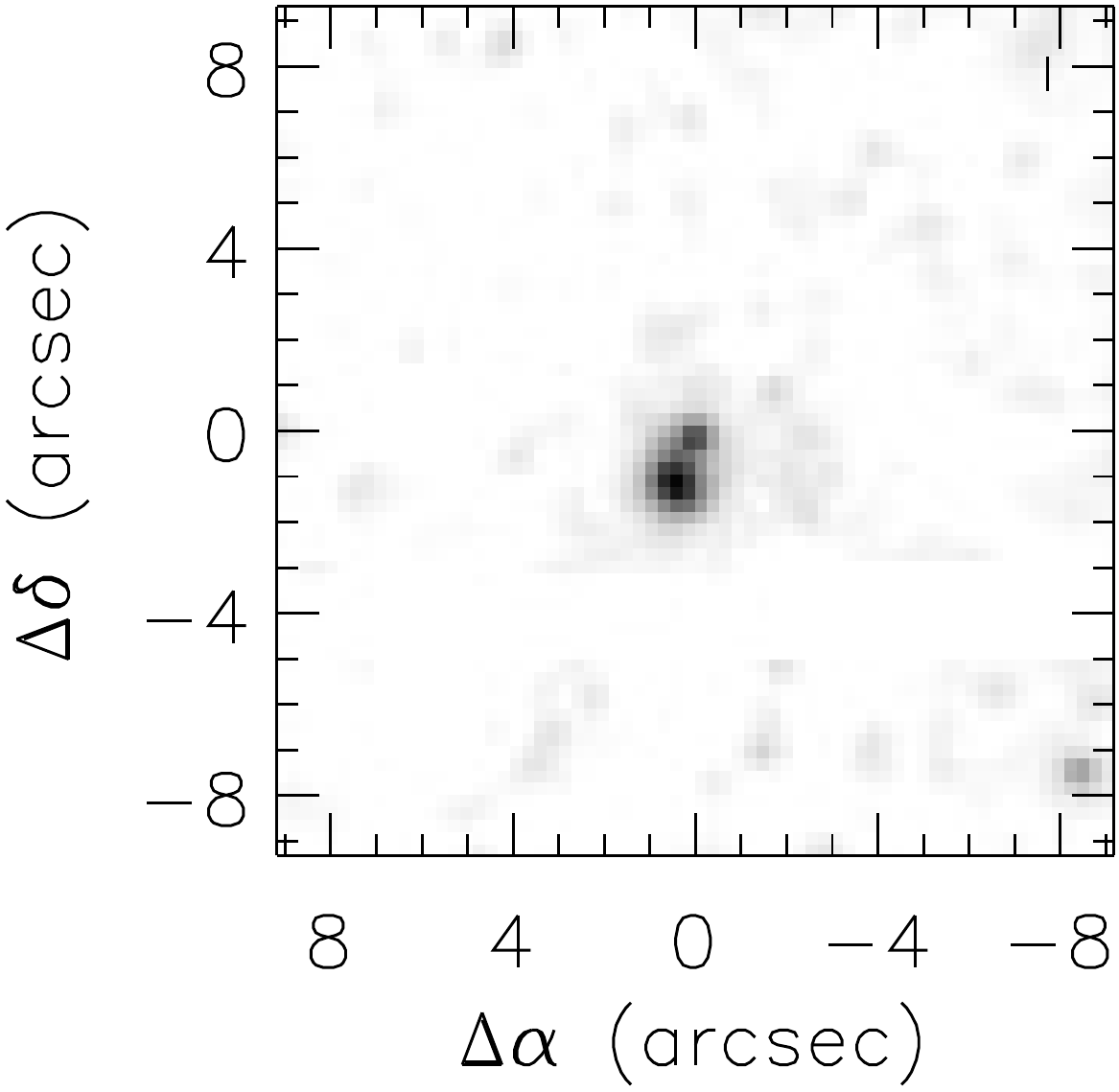}
		\includegraphics[trim=80 0 0 0,clip,height=2.75cm]{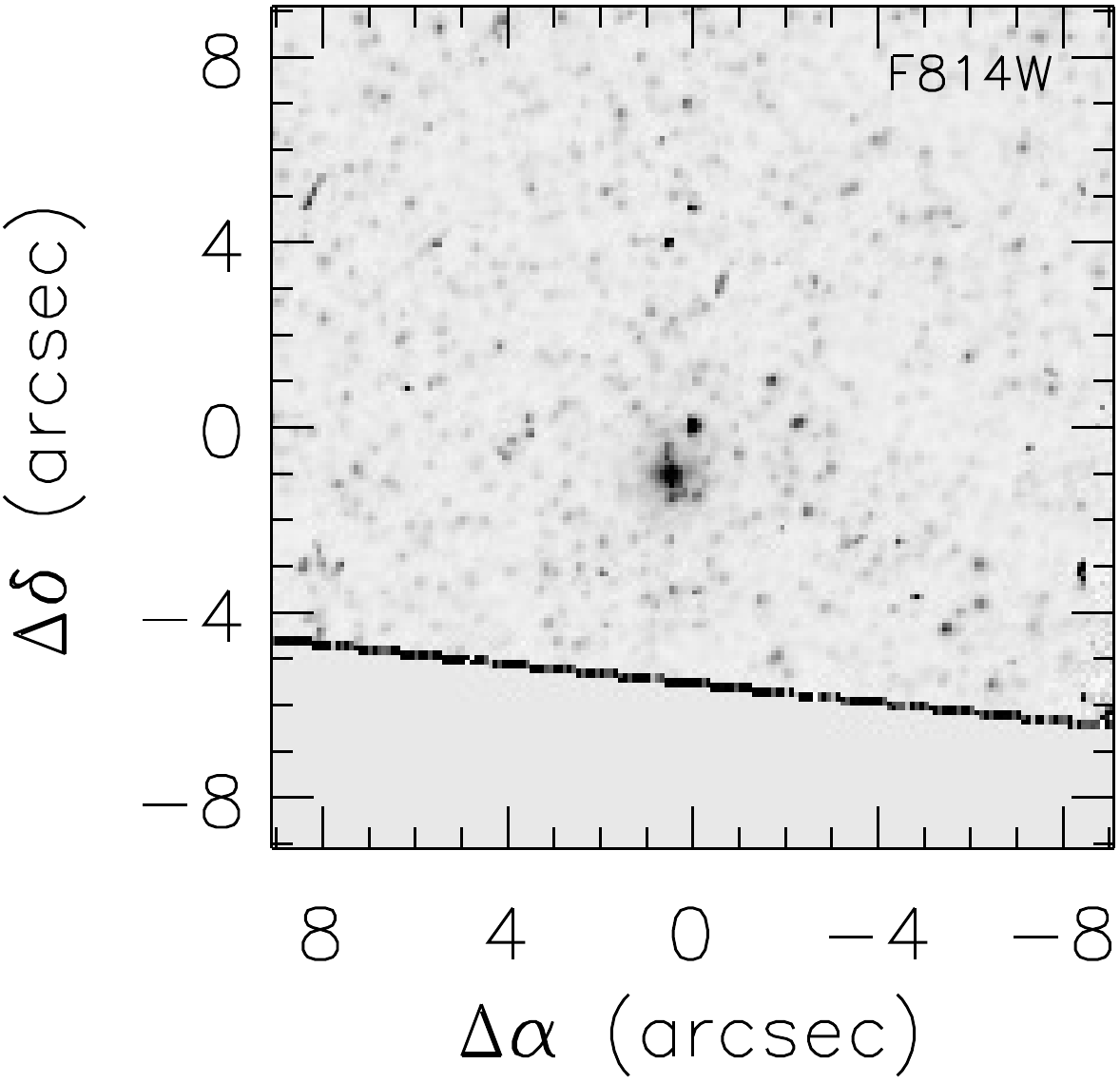}	
	\end{minipage}\hfill
	\begin{minipage}{0.49\textwidth}
		\centering
		\includegraphics[trim=30 10 20  45,clip,angle=270,width=\columnwidth]{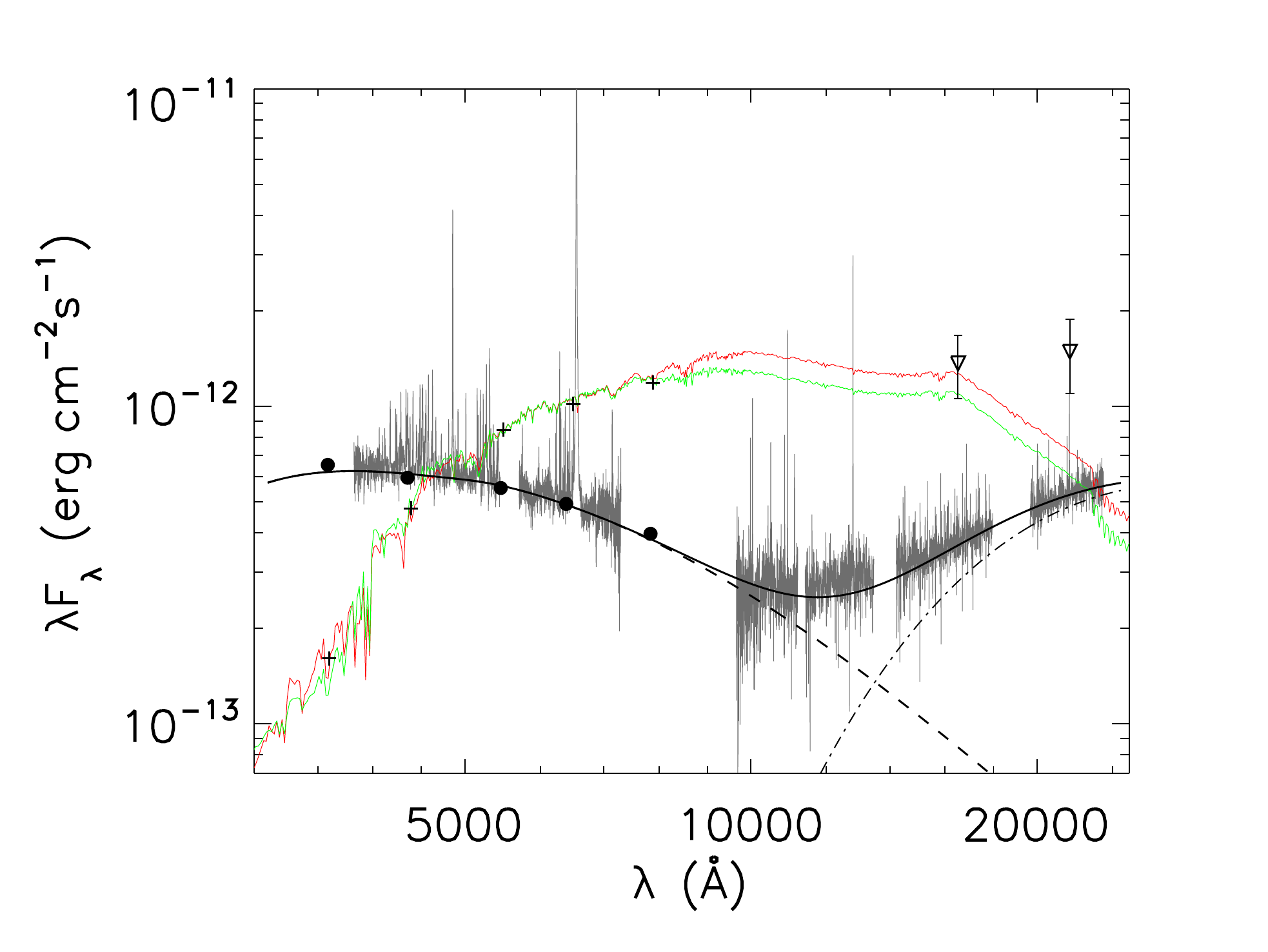} 
	\end{minipage}
\caption{ \textit{Left}: the \textit{U}, \textit{B}, \textit{V}, \textit{R}, \textit{I}, and \textit{F814W}(HST/WFPC2) images of J004229.87+410551.8 and its immediate surroundings. The thumbnail images are 9 arcsec on a side, centered on the object. 
%
\textit{Rigth}: the spectral energy distribution of J004229.87+410551.8 in the optical and NIR ranges. The circles designate the photometric data for the star, the pluses show the SED of the cluster, and the triangles indicate the NIR photometry of the star and cluster without spatial deblending. The dashed line shows the blackbody approximation of the star continuum in the optical part of the SED; the dash-dotted line demonstrates the model of the dust emission contribution in the NIR range; the solid line indicates the total spectrum. Two Starburst99 models that best fit the cluster SED are shown with the red (5 Gyr) and green (900 Myr) lines.}
\label{fig:maps}
\end{figure}

In order to check for possible blending of the star with other sources, archival photometry data were analyzed. From the LGGS data \cite{Massey2006} it has been realized that there is some red source near the star, which stays practically unnoticed compared to the object in the \textit{U} image and is much brighter in the \textit{I} image (see the left panel of Fig.~\ref{fig:maps}). An analysis of the \textit{HST} archive photometry data showed that the red source near the star is a globular cluster.

The fact that there is a cluster near J004229.87 places considerable limits on the angular resolution and seeing of observations. We managed to perform sub-arcsecond optical photometry with the 2.5-m telescope at the Caucasus Mountain Observatory of the Sternberg Astronomical Institute: the \textit{U}, \textit{B}, \textit{V}, \textit{R}, and \textit{I} images of the star with a seeing of $0.6\div0.9$~arcsec were taken on September 13, 2020. The optical spectra of J004229.87 were obtained with the SCORPIO spectrograph on the SAO RAS 6-m telescope (BTA) on September 21, 2020. 
The near-infrared (NIR) spectrum of the star was obtained with the TripleSpec spectrograph on the 3.5-m ARC telescope at the Apache Point Observatory (APO) on October 16, 2012. Quite good seeing during both spectral observations ($\sim\!1$~arcsec) allows us to deblend the star and cluster spectra. We also managed to perform NIR photometry with the NICFPS imager on the 3.5-m ARC telescope at the APO. The \textit{H} and \textit{K}$_s$ images we taken on 26 October 2016, but the atmospheric conditions prevented us from spatial resolving the star and the cluster.

\section{Results}

\begin{figure}
	\includegraphics[width=0.98\columnwidth]{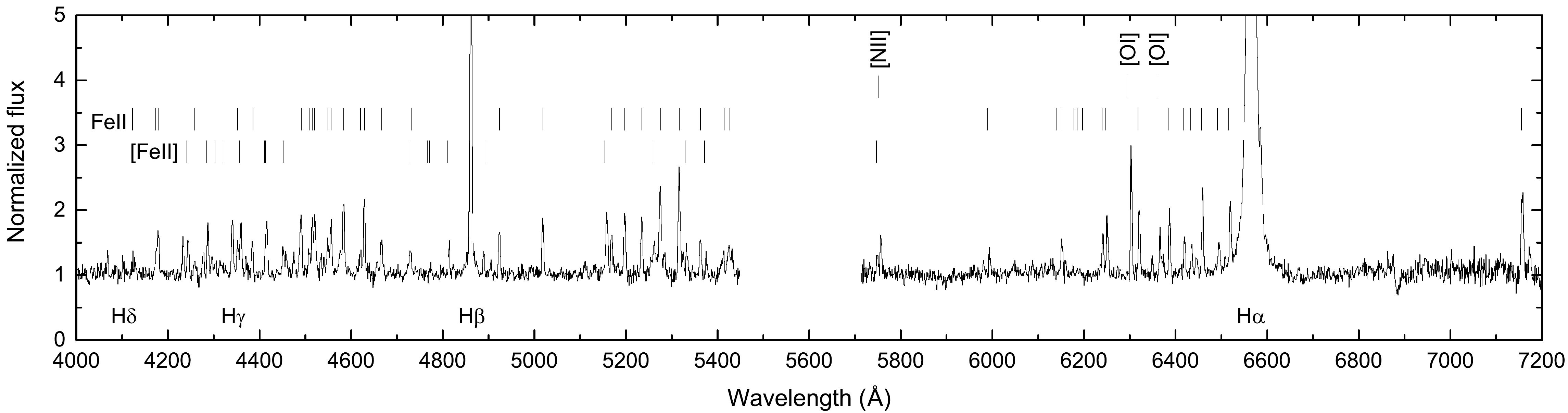} 
	\caption{The optical spectra of J004229.87+410551.8. The main strong lines are identified.}\label{fig:spec}
\end{figure}

Figure~\ref{fig:spec} demonstrates the optical spectra of J004229.87. The star shows the Balmer emission lines, numerous \ion{Fe}{ii} and [\ion{Fe}{ii}] lines, and the [\ion{O}{i}] and [\ion{N}{ii}] lines. At the same time we do not see any prominent absorption lines. The presence of forbidden lines is characteristic of B[e]~stars~\cite{Lamers1998}. Moreover, the [\ion{Fe}{ii}] and [\ion{O}{i}] lines are typical of B[e] supergiants \cite{Aret2016}. 

The spectral energy distribution (SED) of J004229.87 in the optical and near-infrared ranges is shown in 
the right panel of Fig.~\ref{fig:maps}. We clearly see the strong excess in the NIR spectrum of the star due to hot circumstellar dust, which is typical of B[e] supergiants~\cite{Lamers1998}.

To estimate star parameters, we approximated the optical part of the SED with a blackbody spectrum, taking into account interstellar extinction. We also tried to describe the NIR excess due to hot dust emission, assuming it to be  black-body radiation with an effective temperature of $1300$~K. The SED modeling allows us to quite accurately estimate the interstellar extinction $A_V=0.7\pm0.1$ and, therefore, the star's luminosity $\log (L/L_{\odot}) = 4.6\pm 0.2$. We should note that we accepted a distance to M31 of $752\pm27$~kpc \cite{Riess2012} in our calculation. 

We also performed spectroscopy and photometry of the cluster to detect its possible relationship with the star. The absorption lines of the cluster spectrum turn out to be shifted by around $200$~km/s relative to the stellar lines. This speed is two orders of magnitude greater than the escape speed, which was calculated using a projection distance of $\sim\!5$ kpc and a cluster mass $\sim\!104 M_{\odot}$. We also constructed the spectral energy distribution of the cluster and fitted it with the Starburst99 models  (see 
the right panel of Fig.~\ref{fig:maps}).
The best fit was achieved with cluster ages of $500$--$1000$~Myr or $\geq\!5$~Gyr. Therewith, the SED could not be fairly described by young cluster models. So both these facts exclude the star connection with the cluster.

\section{Conclusions}

The spectra of J004229.87 has the \ion{Fe}{ii}, [\ion{Fe}{ii}], [\ion{O}{i}], and Balmer emission lines. Its SED clearly shows an excess in the NIR range due to hot circumstellar dust. The indicated spectral features and the hot dust emission are characteristic of stars with B[e] phenomena \cite{Lamers1998}. This, together with the high luminosity of the star ($\log (L/L_{\odot}) = 4.6\pm 0.2$), allows us to finally classify the object as a B[e]~supergiant.

\section*{Acknowledgments}
Part of the observed data was obtained on the unique scientific facility ``The Big Telescope Alt-Azimuthal'' of SAO RAS, and the data processing was supported under grant 075-15-2022-262 (13.MNPMU.21.0003) of the Ministry of Science and Higher Education of the Russian Federation.

\bibliographystyle{JHEP}
\bibliography{Sarkisyan}

\providecommand{\href}[2]{#2}\begingroup\raggedright\begin{thebibliography}{1}

\bibitem{Massey2007}
P.~{Massey}, R.T.~{McNeill}, K.A.G.~{Olsen}, P.W.~{Hodge}, C.~{Blaha},
  G.H.~{Jacoby} et~al., \emph{{A Survey of Local Group Galaxies Currently
  Forming Stars. III. A Search for Luminous Blue Variables and Other
  H{\ensuremath{\alpha}} Emission-Line Stars}},
  \href{https://doi.org/10.1086/523658}{\emph{\aj} {\bfseries 134} (2007) 2474}
  [\href{https://arxiv.org/abs/0709.1267}{{\ttfamily 0709.1267}}].

\bibitem{Humphreys2014}
R.M.~{Humphreys}, K.~{Weis}, K.~{Davidson}, D.J.~{Bomans} and B.~{Burggraf},
  \emph{{Luminous and Variable Stars in M31 and M33. II. Luminous Blue
  Variables, Candidate LBVs, Fe II Emission Line Stars, and Other
  Supergiants}}, \href{https://doi.org/10.1088/0004-637X/790/1/48}{\emph{\apj}
  {\bfseries 790} (2014) 48} [\href{https://arxiv.org/abs/1407.2259}{{\ttfamily
  1407.2259}}].

\bibitem{Humphreys2013}
R.M.~{Humphreys}, K.~{Davidson}, S.~{Grammer}, N.~{Kneeland}, J.C.~{Martin},
  K.~{Weis} et~al., \emph{{Luminous and Variable Stars in M31 and M33. I. The
  Warm Hypergiants and Post-red Supergiant Evolution}},
  \href{https://doi.org/10.1088/0004-637X/773/1/46}{\emph{\apj} {\bfseries 773}
  (2013) 46} [\href{https://arxiv.org/abs/1305.6051}{{\ttfamily 1305.6051}}].

\bibitem{Humphreys2017a}
R.M.~{Humphreys}, M.S.~{Gordon}, J.C.~{Martin}, K.~{Weis} and D.~{Hahn},
  \emph{{Luminous and Variable Stars in M31 and M33. IV. Luminous Blue
  Variables, Candidate LBVs, B[e] Supergiants, and the Warm Hypergiants: How to
  Tell Them Apart}},
  \href{https://doi.org/10.3847/1538-4357/aa582e}{\emph{\apj} {\bfseries 836}
  (2017) 64} [\href{https://arxiv.org/abs/1611.07986}{{\ttfamily 1611.07986}}].

\bibitem{Massey2006}
P.~{Massey}, K.A.G.~{Olsen}, P.W.~{Hodge}, S.B.~{Strong}, G.H.~{Jacoby},
  W.~{Schlingman} et~al., \emph{{A Survey of Local Group Galaxies Currently
  Forming Stars. I. UBVRI Photometry of Stars in M31 and M33}},
  \href{https://doi.org/10.1086/503256}{\emph{\aj} {\bfseries 131} (2006) 2478}
  [\href{https://arxiv.org/abs/astro-ph/0602128}{{\ttfamily
  astro-ph/0602128}}].

\bibitem{Lamers1998}
H.J.G.L.M.~{Lamers}, F.-J.~{Zickgraf}, D.~{de Winter}, L.~{Houziaux} and
  J.~{Zorec}, \emph{{An improved classification of B[e]-type stars}},
  {\emph{\aap} {\bfseries 340} (1998) 117}.

\bibitem{Aret2016}
A.~{Aret}, M.~{Kraus} and M.~{{\v{S}}lechta}, \emph{{Spectroscopic survey of
  emission-line stars - I. B[e] stars}},
  \href{https://doi.org/10.1093/mnras/stv2758}{\emph{\mnras} {\bfseries 456}
  (2016) 1424}.

\bibitem{Riess2012}
A.G.~{Riess}, J.~{Fliri} and D.~{Valls-Gabaud}, \emph{{Cepheid
  Period-Luminosity Relations in the Near-infrared and the Distance to M31 from
  the Hubble Space Telescope Wide Field Camera 3}},
  \href{https://doi.org/10.1088/0004-637X/745/2/156}{\emph{\apj} {\bfseries
  745} (2012) 156} [\href{https://arxiv.org/abs/1110.3769}{{\ttfamily
  1110.3769}}].

\end{thebibliography}\endgroup

\end{document}